# Large and tunable photo-thermoelectric effect in single-layer MoS$_2$

*Michele Buscema\*, Maria Barkelid, Val Zwiller, Herre S.J. van der Zant , Gary A. Steele\* and Andres Castellanos-Gomez\**

Kavli Institute of Nanoscience, Delft University of Technology, Lorentzweg 1, 2628 CJ Delft, The Netherlands.

M.Buscema@tudelft.nl, G.A.Steele@tudelft.nl, A.CastellanosGomez@tudelft.nl.

We study the photoresponse of single-layer MoS$_2$ field-effect transistors by scanning photocurrent microscopy. We find that, unlike in many other semiconductors, the photocurrent generation in single-layer MoS$_2$ is dominated by the photo-thermoelectric effect and not by the separation of photoexcited electron-hole pairs across the Schottky barriers at the MoS$_2$/electrode interfaces. We observe a large value for the Seebeck coefficient for single-layer MoS$_2$ that, by an external electric field, can be tuned between $-4\cdot10^2$ µV K$^{-1}$ and $-1\cdot10^5$ µV K$^{-1}$. This large and tunable Seebeck coefficient of the single-layer MoS$_2$ paves the way to new applications of this material such as on-chip thermopower generation and waste thermal energy harvesting.

The experimental realization of graphene[1, 2] has opened the door not only to study exciting new phenomena but also to explore a whole new family of two-dimensional materials with complementary properties.[3, 4] For example, atomically thin semiconductor materials with a large bandgap are very interesting for electronic and optoelectronic applications where the lack of bandgap in graphene is hampering its applicability. Single-layer MoS$_2$ presents a large intrinsic bandgap (1.8 eV)[5-9], large in-plane mobility (200-500 cm$^2$V$^{-1}$s$^{-1}$)[6] and remarkable mechanical properties.[10, 11] These properties are of great interest for sensors,[12] flexible circuits[13, 14] and optoelectronic devices.[15-17]

Recent works studying the optoelectronic properties of MoS$_2$ have shown that the photoresponse of externally biased MoS$_2$-based photo-transistors is driven by the change in conductivity upon illumination.[15-17] The photovoltaic effect in MoS$_2$ devices has also been reported with metallic electrodes that generate large Schottky barriers (SBs) and poor electrical contacts.[18, 19] These previous works made use of either an externally applied or a built-in electric field to separate the photogenerated carriers.

Here, we employ scanning photocurrent microscopy to study the photocurrent generation mechanism in single-layer $MoS_2$ transistors with no Schottky barriers and no external bias. We demonstrate that, in contrast to previous studies, the photo-thermoelectric effect dominates the photoresponse in our devices. From our observations, we estimate the electric-field modulation of the Seebeck coefficient for single-layer $MoS_2$, finding a large value that can be tuned by more than two orders of magnitude.

The devices consist of a single-layer $MoS_2$ flake, deposited onto a $Si/SiO_2$ (285 nm) substrate by mechanical exfoliation.[3] Electrical contacts have been fabricated by standard electron-beam lithography and subsequent deposition of a Ti(5 nm)/Au(50 nm) layer. A combination of atomic force microscopy (AFM), Raman spectroscopy and optical microscopy[20] (see. Supp. Info.) has been used to characterize the $MoS_2$ samples. Figure 1(a) shows an AFM image of a field effect transistor (FET) fabricated with a single-layer $MoS_2$ flake. The line trace in Figure 1(b) shows that the height of the flake is around 0.9 nm in agreement with values previously reported in the literature.[20] Raman spectroscopy was also employed to further characterize the deposited $MoS_2$ layers.[21-23] As reported by C. Lee *et al.*[21] the frequency difference between the two most prominent Raman peaks depends monotonically on the thickness and it can therefore be used to accurately determine the number of $MoS_2$ layers. The Raman spectroscopy measurements confirm that our devices are single-layer $MoS_2$ (see Supp. Info.). In order to increase the quality of the electrical contacts, the samples were annealed at 300°C for two hours in a $Ar/H_2$ flow (500 sccm / 100 sccm) [6].

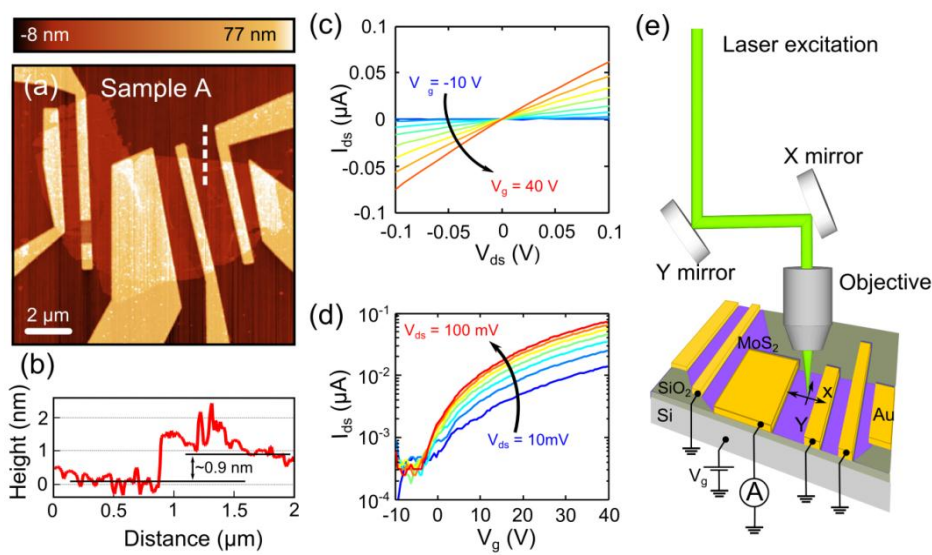

**Figure 1.** (a) Atomic force microscopy image of one of the studied devices, showing the $MoS_2$ flake and the electrodes used to make electrical contact. (b) Line profile over the dashed line in panel (a) showing the $MoS_2$ flake height. (c) Source-drain current *vs.* source-drain bias characteristics measured at different gate voltages. (d) Electrical transport characteristic of a $MoS_2$-based FET device (source-drain current *vs.* gate voltage) measured at different source-drain bias. For both panels (c) and (d) the channel length is 1.6 μm and width is 4.5 μm. (e) Schematic of the scanning photocurrent microscopy (SPCM) setup showing the excitation path and electric circuit used to perform SPCM measurements.

Figure 1(c) shows source-drain characteristics of the single-layer device at different gate voltages shown in Figure 1(a). The current *vs.* voltage relationship remains almost linear for a broad bias range, indicating that the conduction through the device is not dominated by Schottky barriers (SBs), in agreement with previous results by Radisavljevic *et al.* using similar electrode materials and annealing conditions.[6] Moreover, variable temperature transport experiments on $MoS_2$ using low work function electrodes (Ti and Sc) demonstrated a Schottky barrier height slightly larger than $k_BT$ at room temperature[24].

This is also consistent with recent density functional theory simulations and experimental work that demonstrate that $Ti/MoS_2$ interfaces provide ohmic contacts.[25,26] Figure 1(d) shows various gate traces measured at different source-drain voltages. The device shows a pronounced n-type behaviour, with a current on/off ratio exceeding $10^3$ and a mobility of ~ 0.85 $cm^2 V^{-1} s^{-1}$, that is characteristic of single-layer $MoS_2$ FETs fabricated on $SiO_2$ surfaces.[6, 12, 15, 26-28]

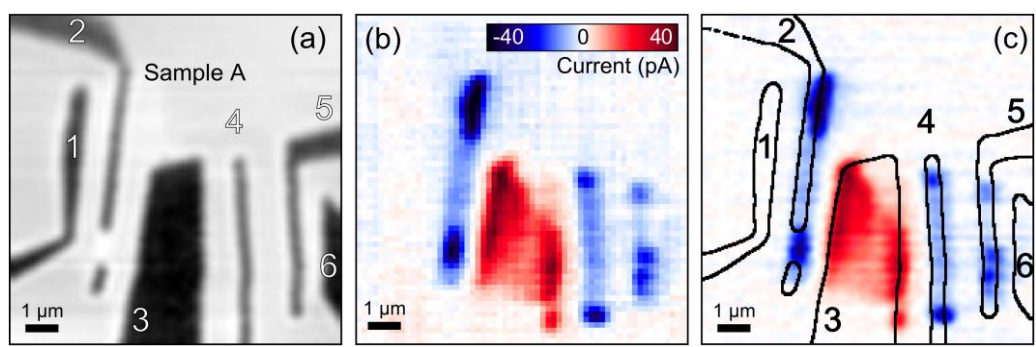

**Figure 2.** (a) Spatial map of the intensity of the reflected light from the device (white corresponds to low reflection). Electrodes have been numbered for clarity. (b) Photocurrent image of the $MoS_2$ FET. The colorscale in the inset gives the photocurrent value. (c) Superposition of the photocurrent map (from (b), same colorscale) and contours of the electrodes as obtained from the light reflection map. The scale bars are always 1 μm. Reflection and photocurrent measurements are performed simultaneously with electrode number 3 connected to a current to voltage amplifier while the other electrodes are connected to ground. Excitation is given by a CW laser, λ = 532 nm, P=1μW, spot waist radius ~400 nm. The region of the flake from the right edge of electrode 2 to the right edge of electrode 3 is composed by multiple layers of $MoS_2$ (see Figure S3 in the Supp. Info. for a more detailed description). No current is seen flowing from electrode 1 or 6 because of poor contact between those with the underlying $MoS_2$ flake.

To spatially resolve the local photoresponse of the $MoS_2$-FET device, we performed scanning photocurrent microscopy (SPCM) in a homebuilt scanning confocal microscope (see Figure 1(e)) with excitation provided by a continuous wave green laser (λ=532 nm) and a supercontinuum tuneable source (at λ=750 nm see Supp. Info. for details).[29] In our experimental setup, the intensity of the reflected laser light (Figure 2(a)) and the photocurrent (Figure 2(b)) generated in the device are simultaneously recorded at every position during the scanning of the laser spot. It is thus possible to superimpose the two images to accurately determine where the photocurrent is generated (Figure 2(c)).

The grey-scale image in Figure 2(c) shows the reflection image and the color-scale shows the simultaneously measured current flowing through the device at zero bias (electrode 3 is connected to a current-to-voltage amplifier while all other electrodes are connected to ground). As it can be seen (e.g. by looking at electrode 3), there is photo-generated current at zero bias even when the laser spot is placed inside the area of the electrodes, microns away from the electrode edges, corresponding to distances up to 10 times larger than the full-width at half-maximum of the laser spot intensity profile. This is in striking contrast with several earlier findings on photocurrent on graphene,[30, 31] which is localized at the interface between the graphene flake and the metal electrodes. In these previous works, the zero-bias photocurrent generation mechanism was attributed to the electron-hole separation at the SBs. This mechanism, however, cannot explain the presence of a photogenerated current when the laser is illuminating the metal electrodes, far from the electrode edges where SBs would be located.

The observation of photocurrent with the laser positioned deep inside the metal electrode suggests that the principal photocurrent generation mechanism in our device is different from photocurrent generation by the separation of photoexcited electron-hole pairs due to the localized electric field at the metal/semiconductor interface. This observation is also consistent with our conclusions from the data in Figure 1(b) that the SBs are not relevant for the transport characteristics of the device.

In order to gain a deeper insight into the photoresponse mechanism of single-layer $MoS_2$ FETs, we have performed scanning photocurrent measurements using different illumination wavelengths. Figure 3(a) shows a photoresponse map acquired with green ($\lambda$ = 532 nm, $h\nu$ = 2.33 eV) illumination. The photovoltage is obtained by dividing the measured photocurrent by the device resistance, measured under the same illumination conditions. A line profile of the photocurrent measured along the dashed green line in Figure 3 (a) is presented in Figure 3(b) as open blue circles. The solid red line is a Gaussian fit to the data corresponding to a diffraction limited laser spot. Notice that there is a significant photocurrent tail generated when the laser is scanned over the electrode (arrow in Figure 3(b)). This is again inconsistent with a response shifted into the $MoS_2$ region, as would be expected for a photovoltaic response from Schottky barriers.

Figure 3(c) shows a photoresponse map acquired with red ($\lambda$ = 750 nm, $h\nu$ = 1.65 eV) illumination. The photovoltage is calculated as in Figure 3(a). Note that the photovoltage under green illumination is larger because of the lower reflectance (and thus higher absorption) of the gold electrodes for this wavelength. The observation of photoresponse even for photon energies lower than the bandgap (Figure 3(c)) cannot be explained by separation of photoexcited electron-hole pairs (see Figure 3(e) – top panel). Moreover, previous photoconductivity measurements in $MoS_2$ transistors under large source-drain bias have not shown any significant photoresponse for excitation wavelength above 700 nm (or below 1.77eV).[15, 32] This indicates that sub-bandgap impurity states are either not present or do not contribute to photocurrents, even in the presence of a large extraction bias, precluding sub-bandgap states as a possible source of photocurrent in Figure 3(c). From these considerations, we conclude that the generation of photocurrent with excitation energies below the bandgap cannot be ascribed to a photovoltaic effect.

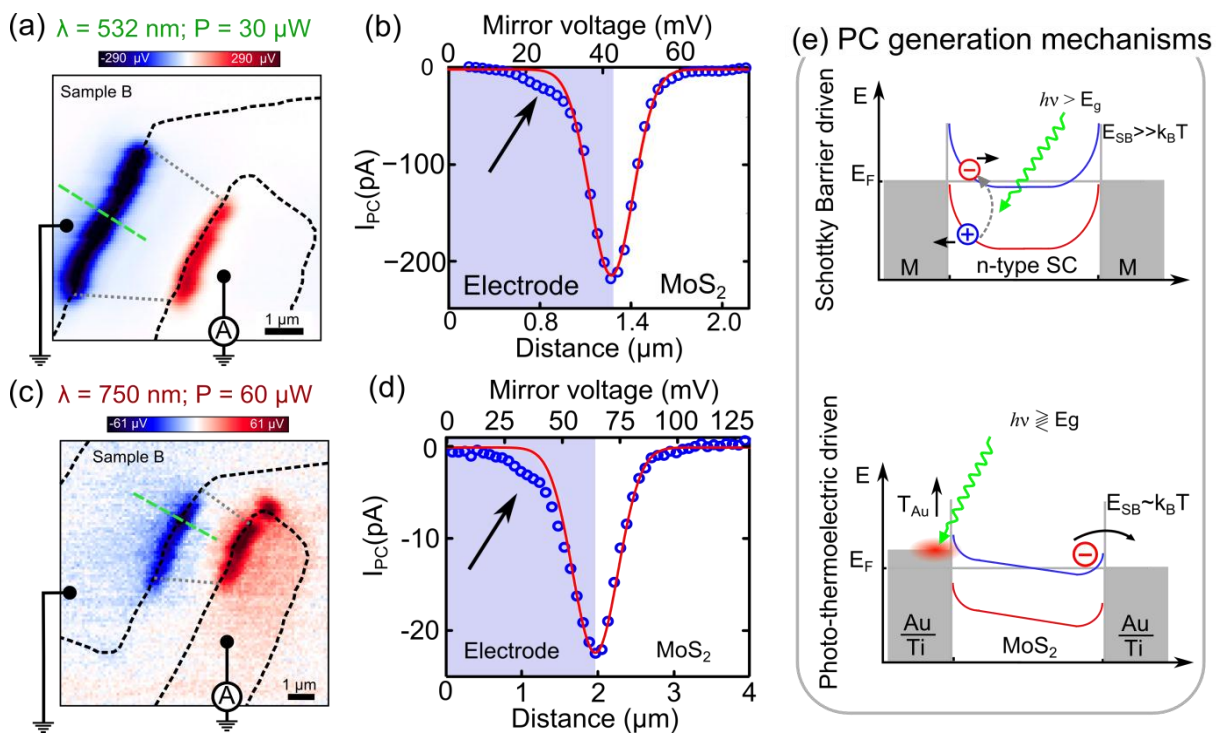

**Figure 3.** Photovoltage map of a single layer MoS$_2$ FET using an excitation wavelength of 532 nm (a) and 750 nm (c). (b)/(d) Photocurrent profile across the linecut in panel (a)/(b) (open blue circles). The solid red line is a Gaussian fit of the data and the arrow points at the photocurrent tail generated when the laser spot is scanned over the electrode. The shaded blue area represents the electrode area as determined by the reflection signal. (d) Schematic of photoresponse mechanism in a typical Metal-Semiconductor-Metal device. (e) Schematic of the photoresponse mechanism in a device dominated by photo-thermoelectric effect. The conduction band is drawn in blue while the valence band is drawn in red.

Interestingly, for above-bandgap illumination, where photovoltaic effects could play a role, the photocurrent images show characteristics that are qualitatively very similar to those for below bandgap illumination. A line profile of the photocurrent measured along the dashed green line in Figure 3(c) is presented in Figure 3(d), in the same fashion as in Figure 3(b) to facilitate the comparison. The qualitative agreement, and in particular the tail of photocurrent when the laser is focussed over the metal, suggests that the photogeneration mechanism is not dominated by photovoltaic effects even for above-bandgap excitation. This hypothesis is also consistent with a lack of the gate dependence of the SPCM images at above bandgap excitation energies (λ = 532 nm). In particular, the position of the maximum photocurrent in the photovoltaic effect would be expected to shift with the gate voltage, an effect also not observed in our devices (see Supp. Info. Figure S5(c)). This also suggests that the main mechanism for photocurrent generation, even with above-bandgap illumination, is not the photovoltaic effect.

The negligible role of the SBs in the conductance of the devices, the strong measured photocurrent inside the area of the electrodes whose position is gate-independent and the observation of qualitatively identical photoresponse between sub and above-bandgap illumination all suggest that photovoltaic effects cannot be responsible for the photoresponse observed here. Instead, we propose that the photoresponse in our device arises from a strong photo-thermoelectric effect.[33-37] In the photo-thermoelectric effect (Figure 3(e) – bottom panel), a temperature gradient arising from light absorption generates a photo-thermal voltage across a junction between two materials with different Seebeck coefficient. This photo-thermovoltage can drive current through the device. This mechanism is consistent with our observation of strong photocurrents when the laser is focussed on the metallic electrodes and also explains the presence of localized and intense photocurrent spots at the edges of the electrodes where the laser absorption is increased and the heat dissipation is reduced. Moreover, it would also explain the stronger photocurrent in the electrode area in Figure 2 where the $MoS_2$ underneath the electrode is more than one layer thick which reduces the thermal coupling with the substrate and thus increases the local temperature of the electrode (see Figure S3 and Figure S8 in the Supp. Info.).

The photo-thermoelectric generation of current can be understood as follows: the local absorption of the laser creates a local heating of the junction between the gold and the $MoS_2$ layer. This local heating of the junction is translated into a voltage difference ($\Delta V_{PTE}$), which will drive current through the device, by the difference between the Seebeck coefficients of the $MoS_2$ flake ($S_{MoS2}$) and the electrodes ($S_{TiAu}$) and the local increase of the junction temperature ($\Delta T$) (See Supp. Info.). This $\Delta T$ can be modelled as the temperature difference of the locally heated gold electrode ($T_{Au}$) and a part of the $MoS_2$ flake which is distant from the junction ($T_{MoS2}$).

$$\Delta V_{PTE} = \left(S_{MoS_2} - S_{TiAu}\right) \cdot \Delta T = \left(S_{MoS_2} - S_{TiAu}\right) \cdot \left(T_{MoS_2} - T_{TiAu}\right) \qquad (1)$$

We have further studied the gate dependence of the photo-thermoelectric effect in single layer $MoS_2$ by measuring the electrical transport characteristics of single $MoS_2$ devices while the laser spot is placed at a location with high photoresponse. For illumination wavelengths with a photon energy higher than the bandgap, the threshold voltage is shifted towards very negative values (see Figure 4(a)), not reachable without leading to gate leakage. It is therefore preferable to use illumination wavelengths with photon energy below the bandgap to minimize this photoconductivity effect.

Figure 4(b) shows the photo-thermoelectric voltage ($\Delta V_{PTE}$) measured for a single-layer $MoS_2$ device at different gate voltages. The photo-thermoelectric voltage is the intercept with the voltage bias ($V_{ds}$) axis of the IV characteristic of the devices measured under sub-bandgap illumination. By decreasing the gate voltage below the threshold voltage, the photo-thermoelectric voltage shows a substantial increase. As the $\Delta V_{PTE}$ is proportional to the difference in the Seebeck coefficients and the temperatures of the AuTi electrodes and $MoS_2$ flake (expression (1)), the observed behavior can be attributed to the expected gate dependence of the Seebeck coefficient of MoS$_2$.

At a microscopic level, the Seebeck effect is due to three microscopic processes which are in dynamic equilibrium with each other.[38, 39] The first process is the diffusion of electrons due to a steady state temperature gradient along a conductor; this process is proportional to the specific heat capacity of the electrons in the conductor. The second process is the variation of the chemical potential with temperature: this will modify the concentration of electrons along the temperature gradient and, therefore, induce a diffusive flux of electrons. The third process is the phonon-drag: as phonons diffuse from the warm side of the conductor to the cold side, they can scatter and drag along electrons. This process is proportional to the electron-phonon coupling and the specific heat of phonons. All these processes depend on or influence the density of states in the conductor. Therefore, they are bound to show a behaviour which is dependent on an external gate electric field (see Supp. Info. for more details).

As it is difficult to experimentally determine microscopic properties such as the energy dependent density of states in a given sample, the Seebeck coefficient is often parameterized in terms of the conductivity of the sample using the Mott relation: [40-43]

$$S = \frac{\pi^2 k_B^2 T}{3e} \frac{d\ln(\sigma(E))}{dE}\bigg|_{E=E_F} \quad (2)$$

where $k_B$ is the Boltzmann constant, $T$ is the temperature, $e$ is the electron charge, $\sigma(E)$ is the conductivity as a function of energy, and the derivative is evaluated at the Fermi energy $E_F$. Using expression (2) we estimate the maximum Seebeck coefficient in our device (see Supp. Info.) to be on the order of -0.2·10$^5$ µV K$^{-1}$, roughly an order of magnitude lower than the value we estimate from our measurements (see below). This discrepancy could arise from the fact that expression (2) is based on the assumption that the conductor is a metal or a degenerate semiconductor, which is not the case for a single flake of MoS$_2$ near depletion.

According to Expression (1) the Seebeck coefficient of MoS$_2$ can be determined from $\Delta V_{PTE}$ and the estimate of the temperature gradient across the AuTi/MoS$_2$ junction. Notice that the Seebeck coefficient of AuTi is negligible with respect to that of MoS$_2$ and no gate dependence is expected. To estimate the increase of temperature induced by the laser illumination, we have performed a finite elements analysis calculation, taking into account the reflections losses in the objective and at the surface of the sample and the absorbed intensity through the material according to $I = I_0(1 - \exp(-\alpha d))$ where $\alpha$ is the absorption coefficient of the material and $d$ is its thickness. With the employed laser excitation ($\lambda$ = 750 nm, $I_0$ = 60 µW and $R_{spot}$ ≈ 500 nm), if we assume that all the energy delivered by the laser is converted into heat, we obtain $\Delta T_{AuTi/MoS2}$ ≈ 0.13 K (see Supp. Info.). Figure 4(c) shows the estimated values: the experimental data are represented by squares while the shaded area represents the uncertainty in the estimation of the Seebeck coefficient deriving from an assumed uncertainty of a factor of 3 in the calculation of the temperature gradient (that is $\Delta T_{AuTi/MoS2}$ in between 0.04 K and 0.4 K). The Seebeck coefficient value for bulk MoS$_2$ (between − 500 µV K$^{-1}$ and − 700 µV K$^{-1}$) is also plotted to facilitate the comparison.[44] A negative value of the Seebeck coefficient is expected for n-type semiconductors.

The obtained value for the Seebeck coefficient is remarkably large and varies strongly with the gate voltage from − 2 10$^2$ ~ - 1.5 10$^3$ µV/K at high doping levels (high positive gate) to -3 10$^4$ ~ - 3 10$^5$ µV/K at low doping levels (high negative gate). Notice that at negative gate voltages, the Seebeck coefficient value saturates. This is due to the large resistance of the device at negative gate voltages which yields to a measured photocurrent that can be below the noise floor of the current amplifier (leading to the observed saturation in the estimated Seebeck coefficient).

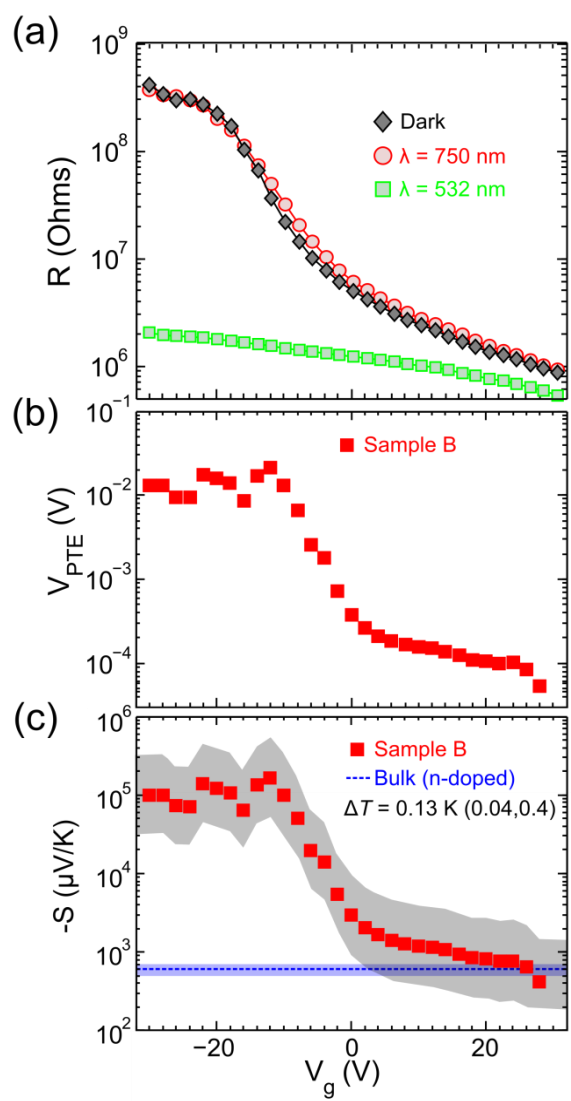

**Figure 4.** (a) Resistance of a single-layer MoS$_2$ device as a function of gate voltage in dark state and with the laser spot placed on the MoS$_2$/electrode interface. Two different illumination wavelengths have been used (532 nm and 750 nm). (b) Photo-thermoelectric voltage for a single-layer MoS$_2$ device measured with the laser spot (λ = 750 nm) placed on the MoS$_2$/electrode interface. (c) Estimated Seebeck coefficient *vs*. gate voltage. The values are calculated from eq. (1) using the measured photovoltage (symbols). The gray shaded area is the uncertainty due to the uncertainty in the estimation of the temperature gradient. The dashed light blue line corresponds to the Seebeck coefficient value of bulk MoS$_2$ with experimental uncertainty (shaded light blue area). The saturation effect at negative gate values is due to the high resistance of the device, leading to a current value below the noise floor of the current-to-voltage amplifier.

The Seebeck coefficient that we observe for single-layer MoS$_2$ is orders-of-magnitude larger than that of graphene (±4 to 100 µV K$^{-1}$)[33, 34, 45-47], semiconducting carbon nanotubes (~ -300 µV K$^{-1}$)[48], organic semiconductors (~ 1 ·10$^3$ µV K$^{-1}$)[49] and even than materials regularly employed for thermopower generation such as Bi$_2$Te$_3$ (±150 µV K$^{-1}$)[50,51]. On the other hand the presented Seebeck coefficient is comparable with other materials such as MnO$_2$ powder (20·10$^3$ ~ 40·10$^3$ μV K$^{-1}$)[52] and Fe$_2$O$_3$ (± 10·10$^3$ μV K$^{-1}$).[53] In addition, the wide gate tunability of the Seebeck coefficient can be useful for applications such as on-chip power generation and thermoelectric nanodevices. These applications include, but are not limited to, energy harvesting of waste thermal energy of other processes,[54,55] and to the possibility of developing autonomously powered devices.[56,57] In all these applications, the tunability of the Seebeck coefficient represents an efficient way of optimizing device performances.[56,57]

In summary, using sub-bandap illumination, we have demonstrated a clear and strong photo-thermoelectric effect arising from the large mismatch between the Seebeck coefficients of the MoS$_2$ and of the electrodes. Furthermore, the identical qualitative characteristics of the photocurrent images for above bandgap illumination suggests that the photothermal effect is also the dominant mechanism for photocurrent generation here as well, and that photovoltaic effects seem not to play a significant role. We estimated the Seebeck coefficient for single-layer MoS$_2$, finding a large value which can be tuned by an external electric field between ~ -4·10$^2$ μV K$^{-1}$ and ~ -1·10$^5$ μV K$^{-1}$. This value is 70-25000 times larger than the values reported for graphene. From these results, we expect that single-layer MoS$_2$ could be an interesting complement to graphene in applications requiring a material with a large and tunable Seebeck coefficient, such as thermoelectric nanodevices or energy harvesting.

ACKNOWLEDGMENT

This work was supported by the European Union (FP7) through the program RODIN and the Dutch organization for Fundamental Research on Matter (FOM).

Supporting information for

# Large and tunable photo-thermoelectric effect in single-layer MoS$_2$ phototransistor

*Michele Buscema\*, Maria Barkelid, Val Zwiller, Herre S.J. van der Zant , Gary A. Steele\* and Andres Castellanos-Gomez\**

Kavli Institute of Nanoscience, Delft University of Technology, Lorentzweg 1, 2628 CJ Delft, The Netherlands.

**Content:**

- **Materials and Methods**
- **Raman spectroscopy characterization**
- **Optical microscopy**
- **Relationship between morphology and SPCM measurements**
- **Dark characterization at low gate values**
- **Gate dependent photocurrent spectroscopy**
- **Photocurrent images at different illumination wavelengths**
- **Laser spot determination**
- **Laser induced temperature increase estimation**
- **Laser power dependence**
- **Analysis of the Seebeck effect**
- **Modelling the gate voltage dependence of the Seebeck effect coefficient**
- **Photo-thermoelectric effect in sample C**

**Materials and Methods**

We prepare $MoS_2$ nanosheets on $Si/SiO_2$ substrates by mechanical exfoliation of natural $MoS_2$ (SPI Supplies, 429ML-AB) with blue Nitto tape (*Nitto Denko Co., SPV 224P*). $Si/SiO_2$ wafers with a 285 nm thermally grown $SiO_2$ layer are used to increase the optical contrast of single-layer $MoS_2$ that are located under an optical microscope (*Olympus BX 51* equipped with an *Olympus DP25* digital camera – see Fig S1). The number of layers is determined by Raman and Atomic Force Microscopy (AFM) imaging. The AFM is used in amplitude modulation mode (*Digital Instruments MultiMode III AFM* with standard cantilevers with spring constant of 40 N m$^{-1}$ and tip curvature <10 nm) to measure the topography and to determine the number of layers of $MoS_2$ flakes in regions that were previously selected by their optical contrast.

$MoS_2$-based Field Effect Transistors (FETs) were fabricated by electron-beam lithography (*Vistec, EBPG5000PLUS HR 100*) on the wafers that were previously doubly spun with two PMMA layers of, respectively, 495kDa and 950kDa average molecular mass. Electrodes are evaporated in a high vacuum e-beam evaporator and are composed of 5 nm /50 nm titanium/gold layers. Subsequently, the PMMA/TiAu layer is lifted off in chloroform. To increase the charge carrier mobility in the flakes prior to electrical characterization, the samples were annealed at 300°C for two hours in an $Ar/H_2$ flow (500 sccm / 100 sccm). The basic electronic transport characterization of the $MoS_2$ FETs is performed at room temperature under high vacuum (<10$^{-5}$ mBar) in a *Lakeshore Crygenics* probestation. The scanning photocurrent microscopy is carried out in vacuum at room temperature in a homebuilt confocal microscope setup with an objective with NA=0.8 and a telecentric optical system, equipped with electrically movable mirrors. The excitation was provided by a continuous wave (CW) solid state laser ($\lambda$ = 532nm) and by a supercontinuum white-light source (*Fianium Ltd*) equipped with an acusto-optical tuneable filter (AOTF).

**Raman spectroscopy characterization**

To confirm the number of layers in the selected $MoS_2$ flakes, Raman microscopy is performed with a micro-Raman spectrometer (*Renishaw in via RM 2000*) in backscattering configuration excited with a visible laser ($\lambda$ = 514 nm) as in Ref.[1]

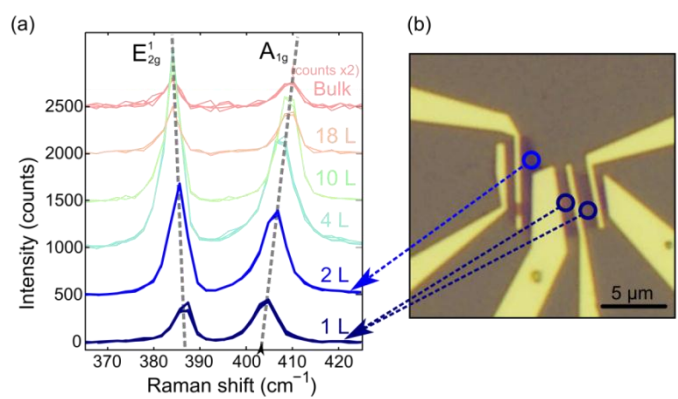

**Figure S1.** Raman and optical characterization of FET $MoS_2$. (a) Raman spectra of various $MoS_2$ flakes deposited on the chip (traces in lighter color have been obtained in other $MoS_2$ flakes, not used to fabricate the devices studied in this work). (b) Optical image of the fabricated FET. The arrows connect the location where the Raman spectra were collected and the actual Raman data.

Figure S1(a) shows a collection of the Raman spectra recorded for different $MoS_2$ flakes deposited on the chip and bulk $MoS_2$. As expected, the difference between the energy of the peaks labelled $E^1_{2g}$ and $A_{1g}$ increases monotonically with thickness of the flake, and, therefore, it has been employed to estimate the number of layers in the flake. Figure S1(b) shows an optical image of one of the FETs characterized in the present work. The arrows point at the Raman spectra recorded at the positions indicated by the blue circles in figure S1(b). The spectra designated by 1L and 2L show a difference in the energies of the $E^1_{2g}$ and $A_{1g}$ peaks of 19 cm$^{-1}$ and 21 cm$^{-1}$ respectively. These values are in agreement with bi-layer and single-layer $MoS_2$ spectra previously reported in literature.[2]

**Optical Microscopy**

Figure S2 shows two optical micrographs of the studied $MoS_2$ FETs. The insets show the $MoS_2$ flakes as deposited on the $SiO_2$/Si substrate. The different contrast can be associated with different thicknesses in the same flake. The Ti/Au electrodes appear homogeneously reflective and this is an indication of good quality of the deposition and smoothness of the metallic surface.

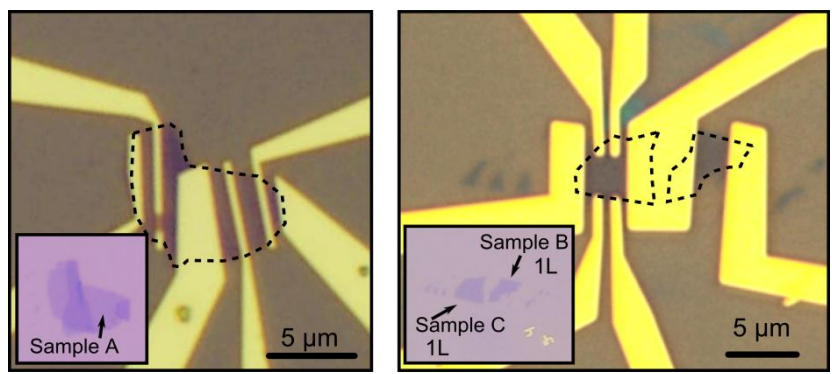

**Figure S2.** Optical micrograph of the 3 single-layer $MoS_2$ devices studied.

**Relationship between morphology and SPCM measurements**

Figure S3 shows the relationship between the morphology of the $MoS_2$ flake and the measured photocurrent. From panel (a) to (d) clockwise, the flake is contoured to show which regions are single layered (1L) or multi layered (>1L), then the electrodes from the optical image in Figure S2 are superimposed to the flake, then the SPCM data from Figure 2(b) are aligned with the electrodes and the flake. Finally the electrodes are removed. This process allowed to precisely determine which regions of the flake correspond to which value of measured PC. It is clearly visible that multilayer regions covered by the electrode have higher PC. This is due to the fact that multilayer $MoS_2$ is thermally decoupled from the substrate, allowing for a higher increase in temperature (see Figure S8) and, thus, a higher current (see Figure 2 and expression 1 in the main text)

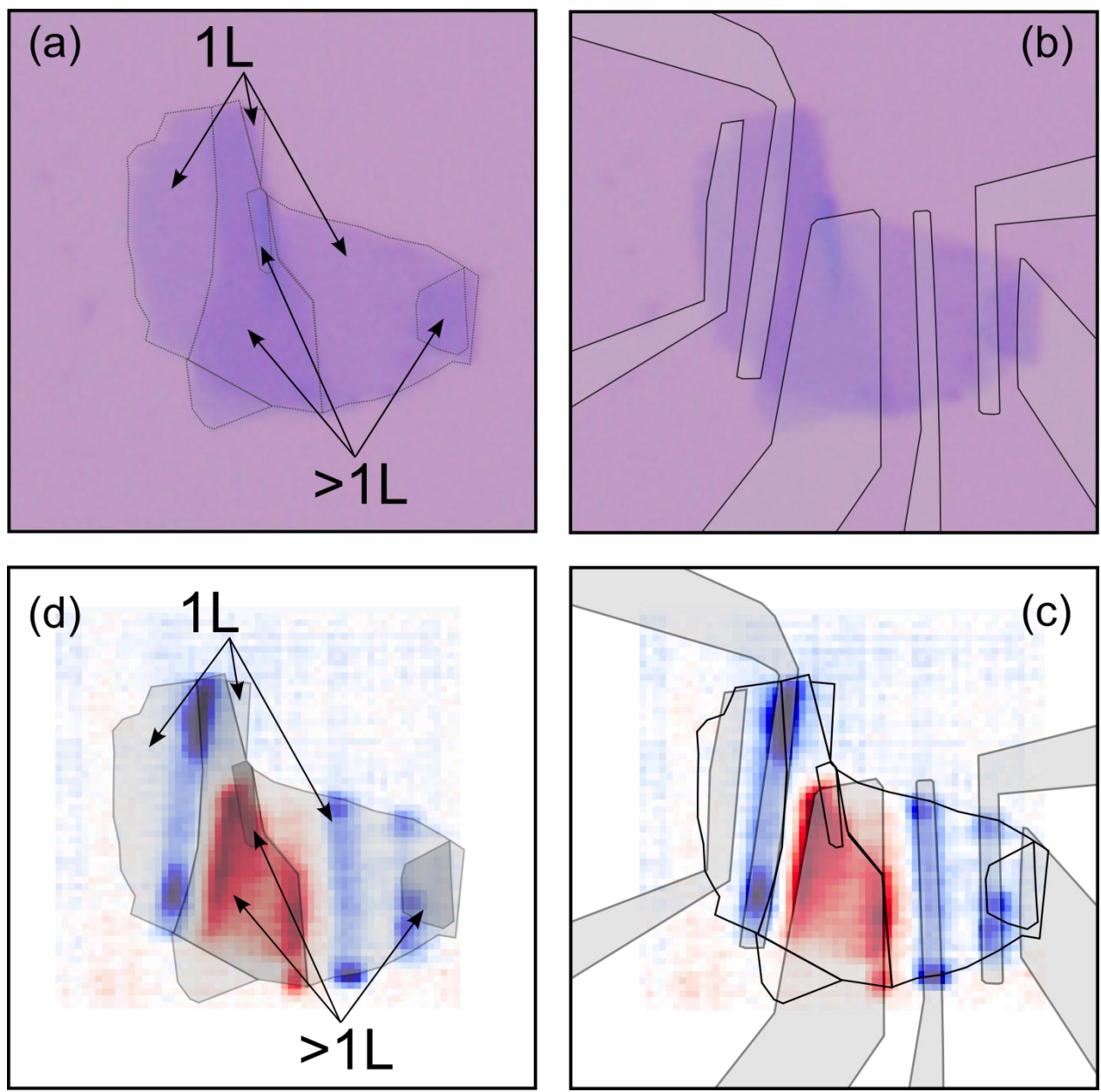

**Figure S3.(a)** MoS$_2$ flake with contours determined from the optical contrast indicating which regions are single layer (1L) or multilayers (>1L). **(b)** MoS$_2$ flake superimposed with the electrodes from Figure S2. **(c)** Data from SPCM superimposed to panel (b). **(d)** MoS$_2$ flake with contours indicating the number of layers and SPCM data obtained by removal of the electrodes.

**Dark characterization at low gate values**

Figure S4 shows drain-source current *vs* drain-source voltage bias characteristics (*IV*s) measured with varying gate voltage from -10 V to 0 V. As it can be clearly seen, the *IV*s are almost linear even up to 300 mV in bias. We can conclude from this that there is Ohmic contact between the Ti/Au electrode and the MoS$_2$ flake, in agreement with previous findings (ref 6 and 24 in the main text).

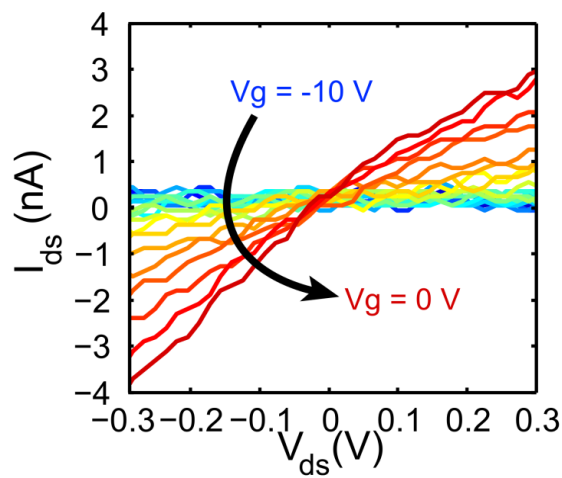

**Figure S4.** Dark Current-*vs*-voltage characteristic for a MoS$_2$ based FET recorded with varying gate voltage from -10 to 0 V.

**Gate dependent photocurrent spectroscopy.**

Figure S5(a) and (b) show a scanning photocurrent microscopy (SPCM) image (at zero source-drain bias) of one of the studied devices and a schematic of the electrical circuit used to measure the photocurrent, respectively.

By performing SPCM measurements with gate voltages ranging from +30V to -30V, displayed in Figure S5(c), we determine that the position of the photocurrent peaks does not change (within experimental resolution) with respect to gate voltage. This is in contrast with earlier findings on devices in which the photocurrent is generated by the separation of photo-excited electrons and holes due to the electrostatic potential in p-n junctions or SBs whose extension depends on the charge carrier concentration. As the charge carrier density is reduced, the potential gradient would extend deeper into the material yielding a maximum photocurrent further and further away from the electrode edges (see Figure S5(d)).[3-8] Therefore, the absence of shift in the position of the measured photocurrent (see Figure S5(c)) demonstrates that the mechanism behind the generated photocurrent is not the separation of carriers by the built-in electrical potential along the SB near the electrodes and supports the proposed mechanism based on the photo-thermoelectric effect. As shown in Figure S5(e), the photo-induced temperature raise in the electrodes generates a voltage ( $\Delta V_{PTE}$, see Expression (1) in the main text) that can drive current through the device. This happens only when the laser is focussed onto the electrode surface as the absorption of laser light by the MoS$_2$ flake is negligible.

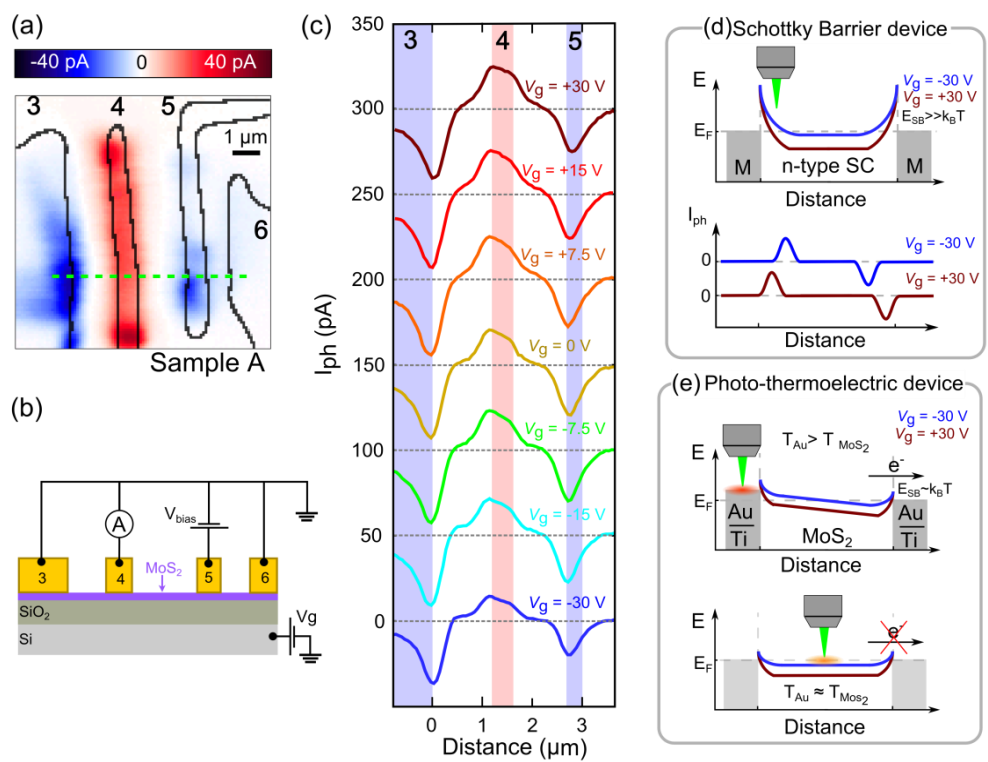

**Figure S5.** (a) Scanning photocurrent image of a single-layer MoS$_2$ FET taken at λ = 532 nm. Note that the current-to-voltage amplifier is connected to electrode 4, while electrode 3, 5 and 6 are connected to ground. (b) Schematic cross-section of the device, showing the electrical circuit used to perform the measurements. (c) Photocurrent line profiles (along the green dashed line in panel (a), λ = 532 nm) for gate voltages of +30, +15, +7.5, 0, -7.5, -15, -30 V from top to bottom. Consecutive lines are shifted by 50 pA for clarity. Shaded areas indicate the position and size of electrodes number 3, 4 and 5 as determined from the reflection signal; the colors of the shading indicates the sign of the photocurrent. (d) Schematic of the energy diagram of a typical Metal-Semiconductor-Metal device (top) and of the magnitude of a photocurrent line profile over such a device showing the expected gate-induced spatial shift in the photocurrent peak (bottom). (e) Schematic of the energy diagram of the single-layer MoS$_2$ device when the laser spot is incident on the electrode (top) and on the MoS$_2$ flake (bottom). In both panels (d) and (e) the conduction band is drawn at different gate voltages and the valence band is not shown for clarity.

**Photocurrent images at different illumination wavelengths**

Figure S6 shows the SPCM maps measured for device B at different illumination wavelengths including both photon energies higher and lower than the single-layer MoS$_2$ gap (1.8 eV).

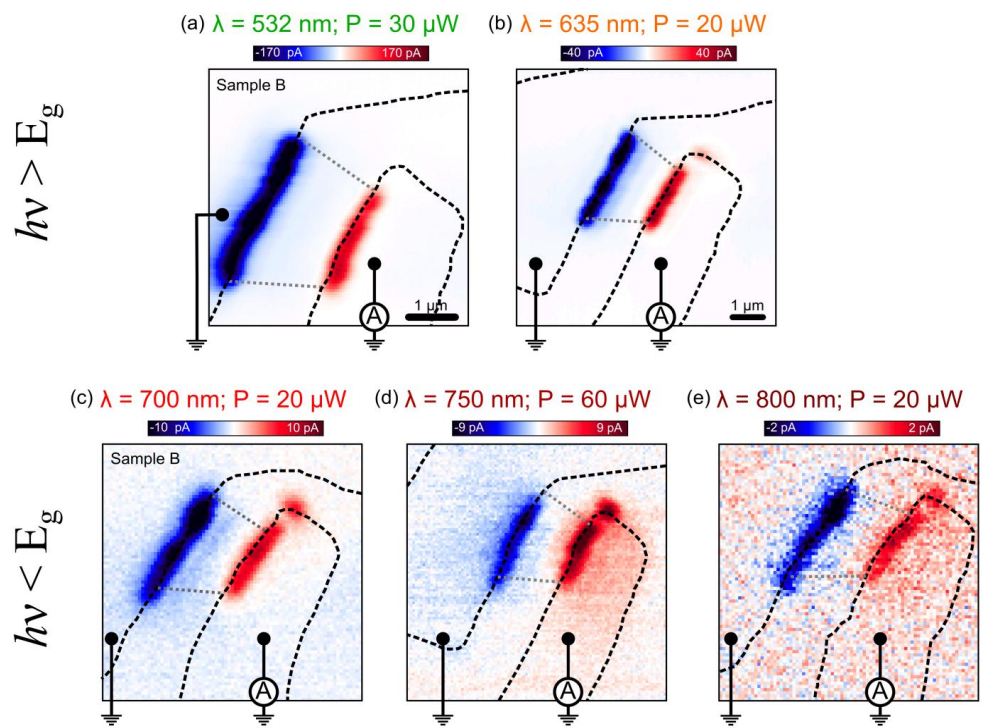

**Figure S6.** Scanning photocurrent images acquired from sample B at different illumination wavelengths with photon energy higher than the band-gap (a) and (b) and lower than the band-gap (c) to (e).

**Laser spot determination**

To estimate the laser spot size, we use the acquired reflectance signal when the laser spot is scanned from the SiO$_2$ over the edge of one of the electrodes. This signal can be interpreted as the convolution of the laser intensity profile and the step edge of the metallic contact. Therefore, by taking the numerical derivative of the signal, we recover the intensity profile of the laser. We performed these operations with the data from measurements performed at excitation wavelength of 532 nm (Figure S7(a)). Figure S7(b) shows the reflection data along a linecut (dashed white line in Figure S6(a)). By fitting the numerical derivative of the reflection signal to a Gaussian, we obtain a full width half maximum (FWHM) of ~ 330 nm, which corresponds to a beam 1/e$^2$ radius of 208 nm (Figure S7(c)). This values are in good agreement with a diffraction limited spot 1/e$^2$ radius of 212 nm calculated with the following formula $r_{spot} = \frac{2\lambda}{\pi}\frac{F}{D} \approx \frac{\lambda}{\pi NA}$ where $\lambda$ is the excitation wavelength and $\frac{F}{D} \approx \frac{1}{2NA}$ is the ratio between the focal length (*F*) and the diameter of the aperture (*D*) of the objective lens that can be approximated to be the reciprocal of twice the numerical aperture (*NA*) of the lens. The small discrepancy can be explained by the very low intensity in the tails of the spot, which would result in a low reflection signal which cannot be measured by the photodiode used to acquire the signal. This mechanism would effectively lead to a smaller spot size in the reflection image.

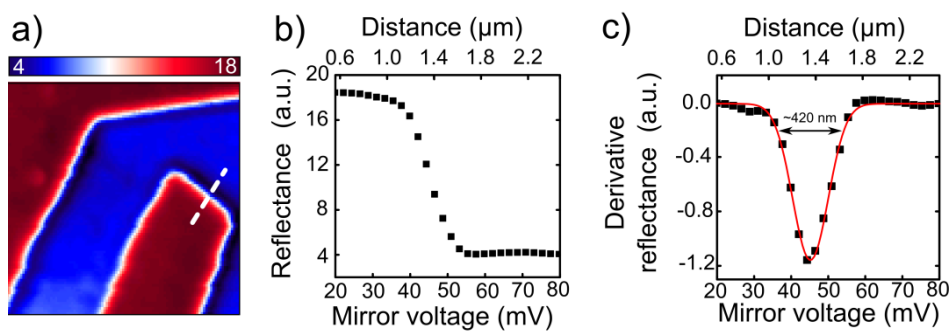

**Figure S7.** (a) Reflection image data. (λ = 532 nm, P = 60 µW). (b) Reflectance of the surface as a function of the mirror voltage along the dotted line in panel (a). (c) Numerical derivative of the data in panel (b) shown as black squares fitted by a Gaussian (solid red line). The relationship between the voltage applied to the mirror and the position of the beam is known.

**Laser-induced temperature increase estimation**

To calculate the Seebeck coefficient through expression (1) in the main text, it is necessary to estimate the temperature difference between the Au electrode and the MoS$_2$ flake induced by the laser heating. The heat exchange process was simulated using a commercially available Finite Element Method (FEM) software package. The system is modelled as presented (out of scale for clarity) in Figure S8(a): a 2D model representative of the cross section of the single-layer MoS$_2$ FET which is symmetric around the axis of the incident laser beam ($z$ axis). This solution allows to increase the number of mesh elements, yielding a more precise solution, while ensuring a reasonable solution time. Note that the roughness of the electrode, as determined with Atomic Force Microscopy is around 5nm, much smaller than the excitation wavelength, and therefore does not play a role in the laser absorption by the electrodes. The material properties used for the simulation are listed in Table S1. The incident laser power density is calculated from the measured power assuming a diffraction limited spot ($R_{spot}$ ~ 500nm for λ = 750nm), a 68% transmission through the objective and the absorption through the thickness of the material according to $I = I_0(1 - \exp(-\alpha d))$ where $\alpha$ is the absorption coefficient of the Au/Ti electrode and $d$ is its thickness (α ~ 5.5 · 10$^5$ cm$^{-1}$ and d = 55 nm). We assume that all the incident laser power is converted into heat. The boundary conditions are as follows: symmetry around the $z$ axis ($r = 0$), incoming heat flux density on the area of the spot size equal to the optical power density calculated above, free heat outlet for the SiO$_2$ and Si free boundaries (as their in-plane size is much bigger than the one of MoS$_2$ and electrodes), room temperature fixed at the bottom of the silicon chip. This is reasonable since the bottom is in good thermal contact with the chip carrier and the cryostat that can be modelled as a thermal bath at constant temperature due to the much bigger mass than the measured sample. Radiation from the other surfaces is neglected. The calculations are performed in the time domain and the results are presented in Figure S8(b) for the two points highlighted by the green circles in Figure S8(a): one on top of the gold electrode and one on top of the MoS$_2$ layer far from the gold electrode. As it can be seen, after 10 ms the system is in steady state and the temperature difference between the electrode and the MoS$_2$ flake is around 0.13 K. The time to reach steady state is lower than the laser spot dwell time during the SPCM measurements, therefore the system can be considered to be in steady state when the measurements are performed. Figure S8(c) shows the result of FEM simulations of the temperature difference between the top of the Au electrode and the MoS$_2$ flake as a function of the flake thickness. It can be seen that the temperature difference increases linearly with the flake thickness. Referring to expression (1) in the main text, the potential difference ( $\Delta V_{PTE}$ ) is proportional to the temperature difference. Therefore, a higher temperature difference will induce a larger $\Delta V_{PTE}$. In Figure 2(b,c) (in the main text), the MoS$_2$ underneath the electrode number 3 is multi-layered, which would explain the increased measured photocurrent when the laser is scanned over the electrode. Moreover, this can explain why the photocurrent inside electrode 3 in Figure 2(c) is larger than the one measured in Figure 3(a,c) where the flake is a single layer.

The calculation at zero thickness is performed without the MoS$_2$ flake, therefore allowing direct contact between the Au/Ti electrode and the SiO$_2$ substrate. The temperature difference is calculated between the top of the Au electrode and a point on the surface of the SiO$_2$ at the same distance of the edge of the flake. The temperature difference in this configuration is around 2mK lower, indicating the thermal sinking effect of the SiO$_2$ substrate.

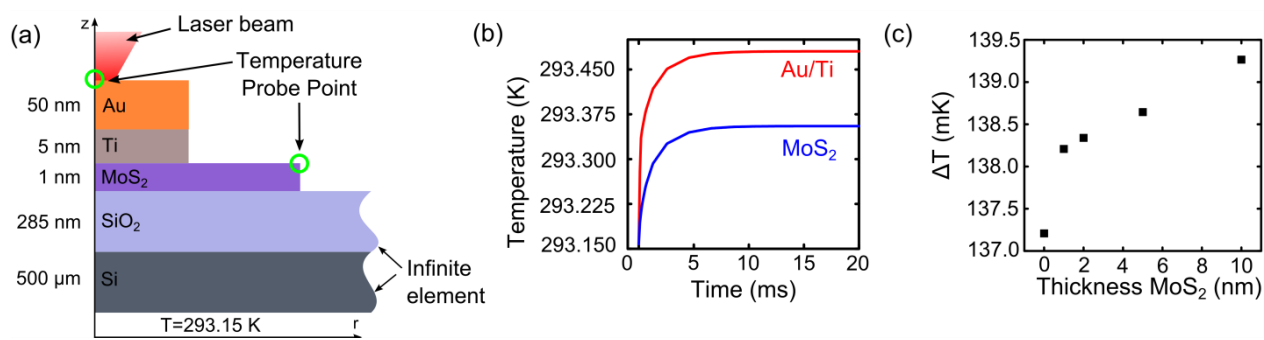

**Figure S8.** (a) Schematic of the modelled system. The system is a 2D cross section of the single layer MoS$_2$ device and is symmetric around the *z* axis which is the centre of the incoming laser beam. The drawing is out of scale for clarity. (b) Results of the FEM simulation in the time domain for two points: one over the Au electrode, one over the MoS$_2$ flake. (c) Difference in temperature between the Au electrode and the MoS$_2$ flake in steady state as a function of the MoS$_2$ flake thickness.

**Laser power dependence**

We performed measurement at different values of laser power to investigate the behaviour of the photo thermal voltage $\Delta V_{PTE}$ with respect to the laser power. Figure S8 clearly shows that $\Delta V_{PTE}$ is linear with the laser power. Also the temperature difference, as determined from FEM simulations, is linear with the laser power. This is in agreement with equation (1) (in the main text) that establishes that the photothermovoltage and the temperature difference are directly proportional.

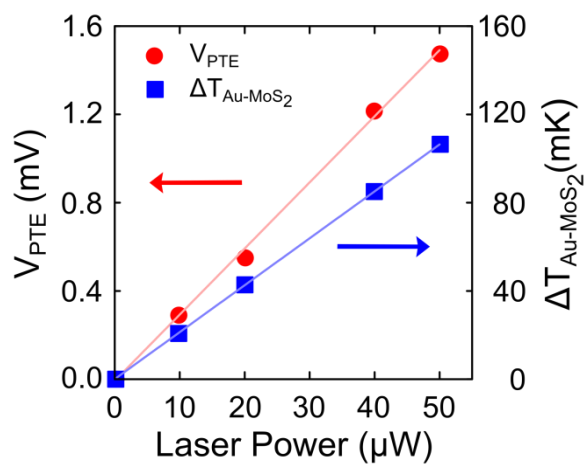

**Figure S9.** Measured photothermoelectric voltage ($\Delta V_{PTE}$) (red dots) and temperature difference between the Au and MoS$_2$ ($\Delta T_{Au-MoS2}$) (blue squares) calculated via FEM simulations as a function of laser power. The lines are calculated least-squares linear fits. Measurements are performed with a sub-bandgap excitation wavelength of 700 nm.

**Analysis of the Seebeck effect**

The Seebeck effect, or thermoelectricity, is due to the coupling between the electrical and thermal transport properties of electrons: when a steady-state temperature gradient is kept across a conductor, a potential difference will develop between the hot and cold ends (Figure S9(a)). The voltage difference and the applied temperature difference are directly proportional and the proportionality constant is the Seebeck coefficient (*S*), or absolute thermopower, of the conductor.

Microscopically, as a temperature gradient is kept on a conductor, this will produce diffusion of electrons from the warm side to the cold side. As electrons carry a charge, this diffusive flux will set up an electric field: this electric field will exert a force on the electrons along the opposite direction and this will balance the diffusive motion induced by the temperature gradient.

This process will give a component of the Seebeck coefficient ($S_e$) which is proportional to the specific heat of electrons.

$$S_e \propto \frac{dE_e}{dT} \propto c_{el}$$

where $E_e$ is the electron energy, *T* is the temperature and $c_{el}$ is the specific heat of the electrons.

The temperature gradient will also induce a concentration gradient. The concentration of electrons can be expressed in terms of their chemical potential (μ); therefore the variation of the electron concentration with temperature can be written in terms of the variation of their chemical potential. This will generate a diffusive flow of electrons in the opposite direction as compared to the previous term.

$$S_\mu = \frac{1}{e}\frac{d\mu}{dT}$$

where *e* is the electron charge.

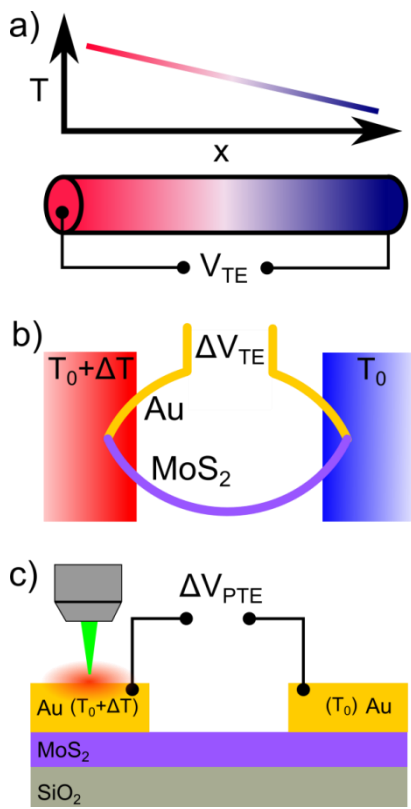

**Figure S10**. (a) Schematic of the system used to derive the absolute thermopower of a conductor. The top panel shows a linear temperature gradient that is applied to a bar of a conductive material, shown in the bottom panel. (b) Thermal circuit representative of the studied $MoS_2$ devices. (c) Schematics of the studied device architecture clarifying the relationship with the thermal circuit in panel (b).

The temperature gradient will also induce diffusion of lattice phonons. If the phonon-electron coupling is strong enough, then phonons can *drag* electrons along, giving rise to a third contribution to the Seebeck effect and coefficient.

$$S_{lattice} \propto \alpha \frac{c_{lattice}}{eN}$$

where α is the electron-phonon coupling, $c_{lattice}$ is the specific heat of the lattice phonons and $N$ is the electron density.

When two conductors are joined in an open circuit and the two junctions are kept at a constant temperature difference, as in Figure S10(b), a voltage will develop over the opening of the circuit. This voltage is proportional to the difference in Seebeck coefficients of the two materials and the temperature difference between the junctions. (Expression 1 in the main text)

$$\Delta V_{TE} = (S_1 - S_2) \cdot \Delta T$$

In the studied $MoS_2$ FET devices, the thermal circuit is built as is in figure S10(c): the hot junction between $MoS_2$ and Au is below the surface of the electrode on which the laser is focused, while the cold junction is represented by the electrode with no laser light on it. Furthermore, the $MoS_2$ is one conductor while the two separated gold electrodes are the open conductor across which the thermoelectric voltage is measured.

**Modelling of the gate voltage dependence of the Seebeck coefficient**

The Seebeck coefficient can be calculated using the Mott relation from the energy dependence of the conductivity[9]:

$$(1) \quad S = \frac{\pi^2 k_B^2 T}{3e} \frac{d\ln(\sigma(E))}{dE}\bigg|_{E=E_F}$$

As the conductivity ($\sigma$) is related to the conductance ($G$) through only geometrical constants, we can also write[10]:

$$(2) \quad S = \frac{\pi^2 k_B^2 T}{3e} \frac{1}{G} \frac{dG}{dE}\bigg|_{E=E_F}$$

Using the gate voltage, we can change the Fermi energy in the $MoS_2$ layer, with a conversion factor (often referred to as the alpha-factor) related to the geometric capacitance. This is a parameter determined by the geometry of the system

$$(3) \quad S = \frac{\pi^2 k_B^2 T}{3e} \frac{1}{G} \frac{dG}{dV_g}\bigg|_{V_g(E=E_F)} \frac{dV_g}{dE}\bigg|_{E=E_F}$$

Following Ayari et al,[11] we estimate $dE/dV_g = 0.01e$ and using the observed gate voltage dependence of the conductance, we obtain the calculated results in Figure S11. Figure S11 shows the experimental results for the Seebeck coefficient and the calculated values using equation (3). As it can be seen the magnitudes are not in agreement. This discrepancy (a factor 5 to 10) could arise from fact that $MoS_2$ is not a degenerate semiconductor, an implicit assumption in the Mott relation (expression (1)). On the other hand, this approximation allows us to reproduce the experimentally observed trend of the Seebeck coefficient.

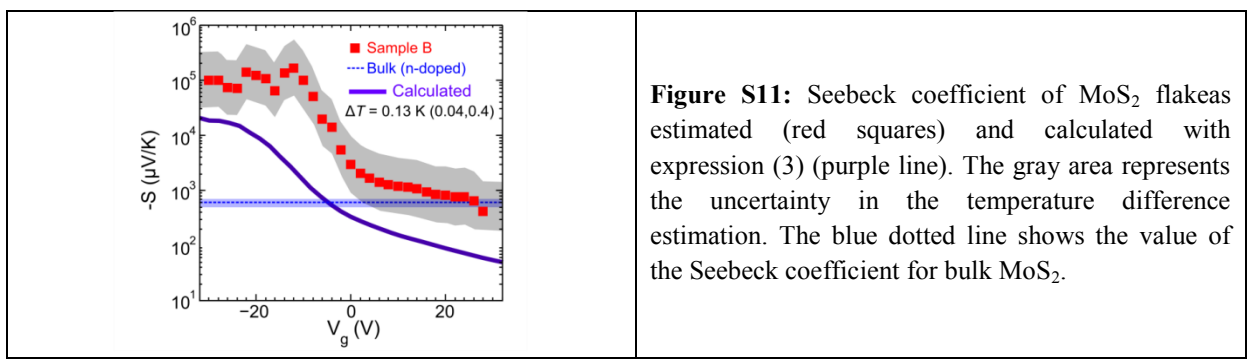

**Figure S11:** Seebeck coefficient of $MoS_2$ flakeas estimated (red squares) and calculated with expression (3) (purple line). The gray area represents the uncertainty in the temperature difference estimation. The blue dotted line shows the value of the Seebeck coefficient for bulk $MoS_2$.

**Photo-thermoelectric voltage in Sample C**

Figure S12(a) shows the measured photo-thermoelectric response of another single-layer MoS$_2$ (Sample C, shown in Figure S2). The gate dependence of the photo-thermoelectric voltage closely resembles that of Sample B (shown in Figure 4(b) in the main text). The magnitude of the $V_{PTE}$, however, in device C is almost a factor of ten lower than in Sample B. We attribute this discrepancy to a lower laser-induce temperature increase in Sample C. In fact, Sample C has been fabricated in a Hall bar geometry and therefore it has many electrodes that can act as heat sinks, reducing the laser-induced temperature increase. Figure S12(b) shows the estimated Seebeck coefficient obtained from measured $V_{PTE}$ and the temperature increase estimated by FEM simulation (see Figure S8).

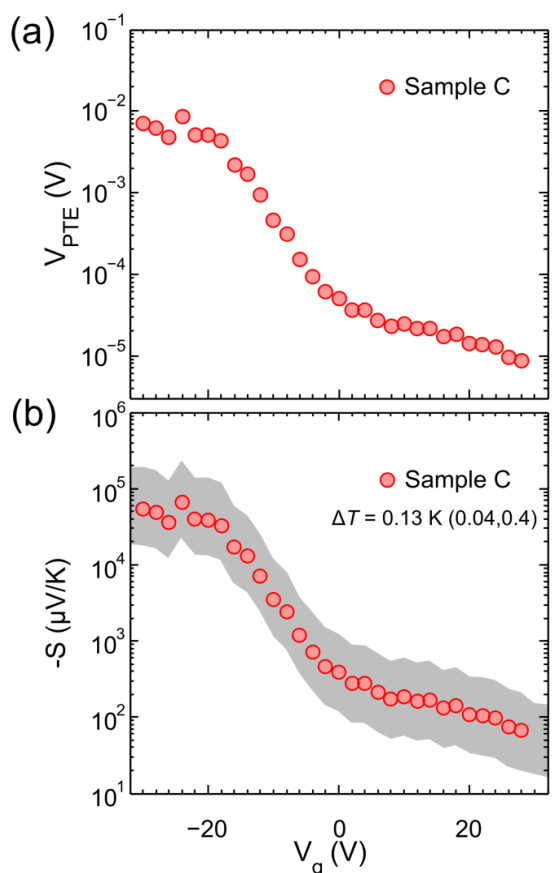

**Figure S12.** (a) Photo-thermoelectric voltage for Sample C measured with the laser spot ($\lambda$ = 750 nm) placed on the MoS$_2$/electrode interface. (b) Estimated Seebeck coefficient *vs*. gate voltage. The values are calculated from Eq. (1) (main text) using the measured photovoltage (symbols). The gray shaded area is the uncertainty due to the uncertainty in the estimation of the temperature gradient.

**Table S1.** Material properties values employed for thermal gradient estimation

|  | Au | Ti | MoS$_2$ | SiO$_2$ | Si |
|---|---|---|---|---|---|
| d (nm) | 50 | 5 | 1 | 285 | 500000 |
| $\rho$ (kg m$^{-3}$) | 19300 | 4500 | 5000 | 2650 | 2330 |
| C (J kg$^{-1}$ K$^{-1}$) | 128.74 | 540 | 400 | 1000 | 710 |
| $\kappa$ (W m$^{-1}$ K$^{-1}$) | 315 | 21.9 | 1.8 | 1.38 | 149 |